\documentclass[twocolumn,aps,showpacs,amsmath,amssymb,preprintnumbers,epsf]{revtex4}
\usepackage{graphicx}
\usepackage{dcolumn}
\usepackage{color}
\usepackage{bm}

\begin{document}

\title{\textbf{Numerical detection of stochastic to deterministic transition}}
\author{R.K. Brojen Singh}
\affiliation{Centre for Interdisciplinary Research in Basic Sciences, Jamia Millia Islamia,New Delhi 110025, India.}
\address{*Corresponding Author at: Centre for Interdisciplinary Research in Basic Sciences, Jamia Millia Islamia,
New Delhi 110025, India,Tel: +911-2698 1717 (4492)
Telfax: +911-2698 3409}
\address {Email:rksingh@jmi.ac.in}
\date{\today}

\begin{abstract}

{\begin{center}\bf ABSTRACT\end{center}}
{We present the numerical estimation of noise parameter induced in the dynamics of the variables by random particle interactions involved in the stochastic chemical oscillator and use it as order parameter to detect the transition from stochastic to deterministic regime. In stochastic regime, this noise parameter is found to be increased as system size decreases, whereas, in deterministic regime it remains constant to minimum value as system size increases. This let the transition from fluctuating to fixed limit cycle oscillation as the system goes from stochastic to deterministic transition. We also numerically estimated the strength of the noise parameter involved both in Chemical Langevin equation and Master equation formalisms and found that strength of this parameter is much smaller in the former than the later.
}
\\\\
\emph KEYWORDS
{: Noise parameter, Chemical oscillator, fluctuating limit cycle oscillation, Master equation, Chemical Langevin equation.}
\end{abstract}
\maketitle

The origin of noise (intrinsic and extrinsic) in stochastic systems is believed to be due to random particle or molecular interactions taking place among the particles or molecules in the physical, chemical or biological systems and various forms of fluctuations in the environment surrounding the system \cite{rao, swa}. The strength of the noise in such complex systems depends on various parameters, for example, system size $V$, population of particles or molecules accomodated in the systems ($N$) and various types of fluctuations in the surrounding environment etc \cite{gil1, gil2}. For instance, the noise strength associated with single cell gene expression scales as $N^{-1}$ of relative fluctuation amplitude \cite{bla}. The estimation of the noise associated with stochastic systems via Master equation formalism could tell us the role of noise in the system dynamics \cite{swa,sco}. For instance, noise become an essential parameter which plays a constructive role in biological systems and is used in weak signal amplification and enhancing the detection of information carrying weak signal, the phenomenon known as stochastic resonance \cite{han,ani}. Further this parameter might also be used by a group of systems as a means of synchronizing or correlating behaviors of each individual systems as is seen in various multicellular organisms \cite{han} and in a group of identical unicellular organisms \cite{bas}. 

The deterministic systems can be considered as noise free systems \cite{gil1, gil3}. Because the noise associated with the dynamics of each variables involved in deterministic systems becomes negligible \cite{gil3}. The transition from stochastic to deterministic system can be scaled by defining a thermodynamics limit, $V\rightarrow\infty$, $N\rightarrow\infty$ but $N/V\rightarrow~finite$ \cite{gil2,gil3}. Attempts have been made in genetic oscillator by calculating noise-to-signal ratio as a function of some reaction rate to differentiate deterministic from stochastic systems \cite{sco1}. However estimation of this limit or to define correct order parameter to distinguish between stochastic and deterministic systems is still an open question both for numerical as well as experimental situations. 

The noise fluctuation is an inherent property of most of the biological systems and has two contrast roles depending upon its magnitude. If the strength of the noise is large enough (above a critical limit, say $\eta\rangle\eta_c$), then it destructs the signal. However if the strength of the noise is weak, say $\eta\langle\eta_c$ then its role becomes constructive and is used to detect and amplify the weak signal (stochastic resonance) \cite{ani, han}. In this work we try to identify a parameter that can separate stochastic and deterministic systems. Probably the study may unfold many interesting roles of noise in various biological systems. For this purpose we study a chemical oscillator defined by an oregonator reaction network via Master equation formalism. Then we briefly explain the stochastic simulation algorithm (SSA) \cite{gil1} to be implemented to simulate the reaction network. Further we study Chemical Langevin formalism of this model to measure noise parameter induced in the system dynamics. It is followed by results and discussions, and then few conclusions based on the results we have got.

The intrinsic noise in stochastic systems is due to random particle or molecular interactions taking place in the system and is an inherent property associated with the stochastic system \cite{gil1}. If we consider a configurational state $\vec x(t)=[x_1,x_2,...,x_N]^{-1}$ consisting of $N$ species (molecule or particle) at any instant of time $t$, then the random interaction events (only collisions which gives rise decay and creation of particle) taking place in the system is given by,
\begin{eqnarray}
S_ax_{a}+S_bx_b+\cdots\stackrel{k_{\mu}}{\rightarrow} S_px_p+S_qx_q+\cdots
\end{eqnarray}
where, $\{S\}$ are co-efficients of the species in the $\mu$th interaction. $k_{\mu}$, with $\mu=1,2,\cdots,M$ is the rate of interaction of the species $\{x_i\}$, $i=a,b,\cdots$ to give species $\{x_j\}$, $j=p,q,\cdots$ in a certain interval of time $[t,t+\Delta t]$. This leads to the change in states from $\vec x(t)$ at time $t$ to a new state $\vec x^{\prime}(t+\Delta t)$ during the time interval. Depending upon the magnitude of $\Delta t$, there could be $L$ number of interactions in series took placed. If we simplify the transitions by taking an infinitesimal time interval $dt$ such that during $[t,t+dt]$ only one interaction is occured then we can write $\Delta t=dt^{[1]}+dt^{[2]},\cdots,+dt^{[L]}$, which leads to a series of microscopic state changes, $\vec x\rightarrow\vec{x}^{[1]}\rightarrow\vec{x}^{[2]}\cdots\rightarrow\vec{x}^{[L]},\vec{x}^\prime$. These $\{dt^{[j]}\}$, $j=1,2,\cdots,L$ are not necessarily the same but taking $dt^{[1]}=dt^{[2]}=\cdots dt^{[L]}=dt$, we have $\Delta t=Ldt$ leading to macroscopic state change $\vec{x}\stackrel{L}{\rightarrow}\vec{x}^\prime$. 
\begin{figure}
\begin{center}
\includegraphics[height=220 pt]{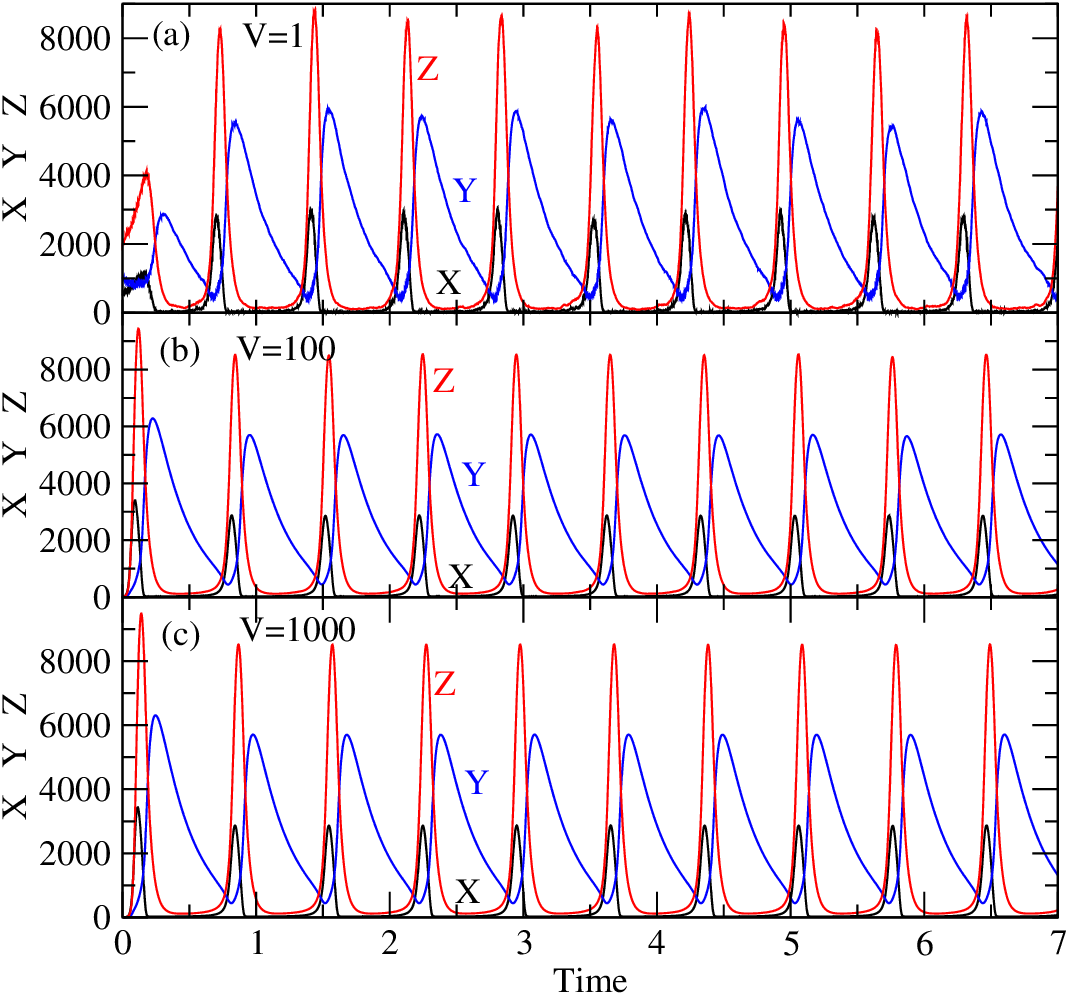}
\caption{The plots of the dynamics of $X$, $Y$ amd $Z$ at three different system sizes, $V=1$, 100 and 1000 respectively simulated using Gillespie's SSA \cite{gil1}. The parameters used are same as is used in Gillespie, (1977) \cite{gil1}.}
\end{center}
\end{figure} 

Now if we consider each microscopic state change during $[t+dt^{[j]},t+dt^{[j^\prime]}]$, $\{j,j^\prime=1,2,\cdots,L;j\ne j^\prime\}$ then the time evolution of the state probability, $P_{j}(\vec{x}_j,t)$ of the state change based on transition probabilitis $\{W\}$ of decay and creation of the particles evolved in the interaction event can be described by the following Master equation \cite{mcq,gil1},
\begin{eqnarray}
\frac{\partial P_{j}(\vec{x}_j,t)}{\partial t}&=&
-\sum_{\vec{x}_{j^\prime}}P_{j}(\vec{x}_j,t)W_{\vec{x}_j\rightarrow \vec{x}_{j^\prime}}\nonumber\\
&&+\sum_{\vec{x}_{j}}P_{j}(\vec{x}_{j^\prime},t)W_{\vec{x}_{j^\prime}\rightarrow \vec{x}_j}
\label{master}
\end{eqnarray}
The macroscopic state change can be described by a series of state probabilities corresponding to each microscopic state changes $P(\vec{x},t)\rightarrow P_1\rightarrow P_2\rightarrow,\cdots,\rightarrow P_L\rightarrow P(\vec{x}^\prime,t)$.
Since each interaction that drives a particular microscopic state change is random in nature, the trajectory of $P_j$ from $P(\vec{x},t)$ to $P(\vec{x}^\prime,t)$ will a Brownian trajectory.
\begin{figure*}
\begin{center}
\includegraphics[height=300 pt]{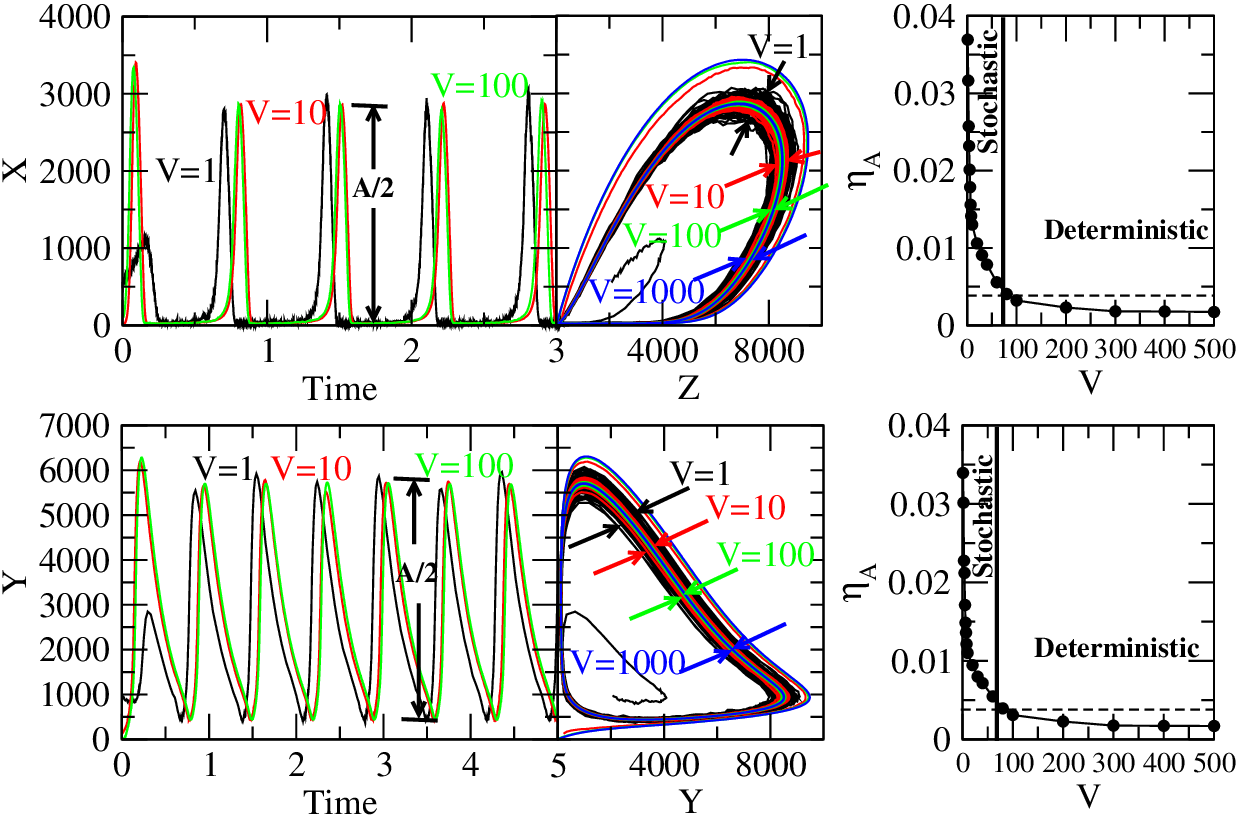}
\caption{The plots showing the estimation of noise parameter in amplitudes for $X$ and $Y$ variables. The left panels show the transition from fluctuating to fixed limit cycle oscillation as the function of $V$, as $V$ goes from small (stochastic regime) to large $V$ (deterministic regime). The right panels show the phase plot in ($\eta_A,V$) plane indicating stochastic and deterministic regimes.}
\end{center}
\end{figure*}

We consider the chemical oscillator model known as oregonator devised by Field and Noyes \cite{fie} based on the criticism made by Tyson and Light \cite{tys} on Brusselator model which is two molecular species reaction model. The modified chemical oscillator model consists of three molecular species, $X$, $Y$ and $Z$ involved in the following five reaction channels,
\begin{eqnarray}
\label{ore}
A+Y&\stackrel{k_{1}}{\rightarrow}& X \nonumber \\
X+Y&\stackrel{k_{2}}{\rightarrow}& B \nonumber \\
C+X&\stackrel{k_{3}}{\rightarrow}& 2X+Z \\
2X&\stackrel{k_{4}}{\rightarrow}& D \nonumber \\
E+Z&\stackrel{k_{1}}{\rightarrow}& Y \nonumber
\end{eqnarray}
where $\{k_i\}$, $i=1,2,\cdots,5$ are reaction rate constants and $A$, $B$, $C$, $D$ and $E$ are constants. The state of the system at any instant of time $t$ is given by the vector, $\vec{x}(t)=[X(t), Y(t), Z(t)]^{-1}$. Considering only microscopic state change, we can construct molecular transformation diagram \cite{mcq} and derive transition probabilities $\{W\}$ of each reaction channel during the time interval $[t,t+dt]$. Depending upon the arrow of the diagram and state change due to molecular decays and creations involved in the molecular interactions indicated by the reactions listed in \ref{ore}, we obtain the following Master equation (ME) of the system,
\begin{widetext}
\begin{eqnarray}
\label{more}
\frac{\partial}{\partial t}P(X,Y,Z;t)&=&k_1A(Y+1)P(X-1,Y+1,Z;t)+k_2(X+1)(Y+1)P(X+1,Y+1,Z;t)\nonumber\\
&&+k_3C(X-1)P(X-1,Y,Z-1)+\frac{1}{2}k_4X(X+1)P(X+1,Y,Z)+k_5E(Z+1)P(X,Y-1,Z+1)\nonumber\\
&&-\left[k_1AY+k_2XY+k_3CX+k_4X^2+k_5EZ\right]P(X,Y,Z)
\end{eqnarray}
\end{widetext}
However, it is very difficult to solve this Master equation (\ref{more}). The alternative way to solve such complicated molecular processes in stochastic system given by this equation is to do simulation by simplifying the process of jumping from one stochastic state to another as discrete Markov process \cite{gil1}. This stochastic simulation algorithm (SSA) due to Gillespie, is based on the assumption that the time interval to jump from one stochastic state to another is taken to be small enough such that at most one reaction type can occur. The algorithm systematically allows to pick up each reaction event randomly from the reaction set at every time step to allow transition from one state to another along the trajectory of the of the variables by defining a joint probability density function, $\Gamma(\tau,\mu)=\Pi(\tau)\Lambda(\mu)$. $\Lambda(\mu)$ is the probability density function of picking up $\mu$th reaction and $\Pi(\tau)$ is the probability density function that at time step $\tau$ the reaction $\mu$ will fire. The reaction time and reaction number fired at that time can be estimated computationally by generating two uniform random numbers $r_1$ and $r_2$ to identify $\tau$ and $\mu$ by $\tau=\frac{1}{\sum_i\omega_i}ln\left[\frac{1}{r_1}\right]$ and $f_\mu=r_2\sum_i\omega_i$ respectively by imposing the condition $\sum_{i=1}^\mu\omega_i\le f_u\langle\sum_{i=1}^{\mu+1}\omega_i$, where $\{\omega\}$ are propensity functions.
\begin{figure*}
\begin{center}
\includegraphics[height=210 pt]{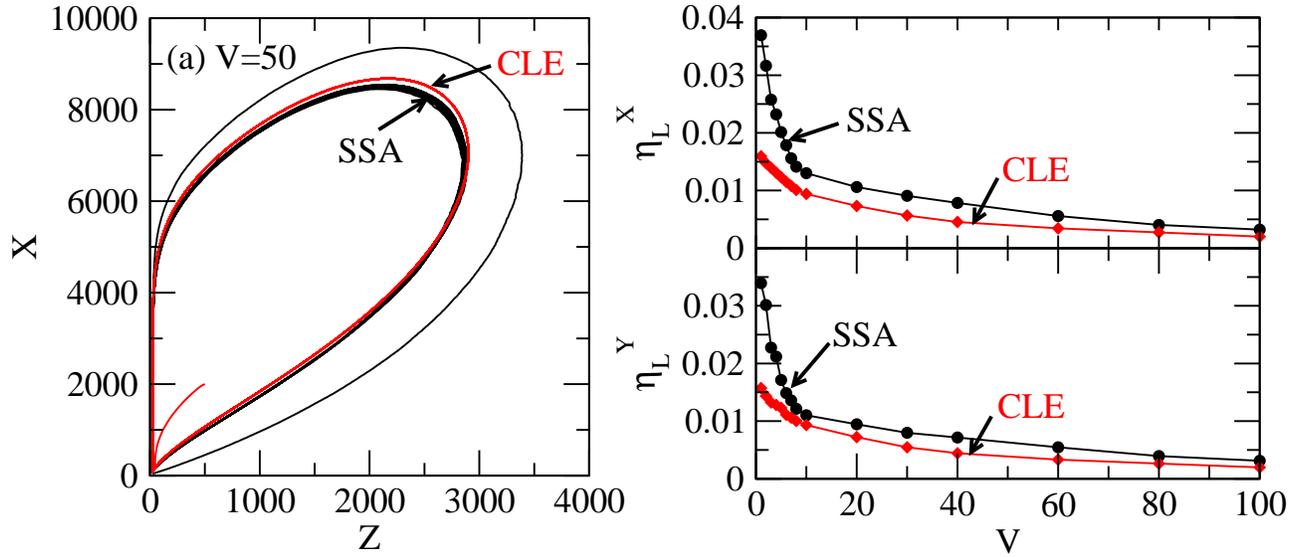}
\caption{The plots of $\eta_L^X$ and $\eta_L^Y$ as a function of $V$ both for CLE by simulating equations (\ref{cle}) and ME by simulating reactions (\ref{ore}) using SSA.}
\end{center}
\label{}
\end{figure*} 

Now we solve the stochastic system described by the oregonator reaction model given by (\ref{ore}) using SSA and the results are shown in Fig.1. In the figure we present the dynamics of X, Y and Z as a function of time for different values of system sizes $V$. It is seen from the dynamics that at small $V$(=1), the noise fluctuation associated with each variable dynamics is large as compared to the behavior at large $V$(=100, 1000). Further the amplitudes of each variable oscillation at small V are random in nature due to noise. The degree of randomness in amplitudes which is proportional to the magnitude of noise is indicated by the width of the fluctuating limit cycle oscillation shown in Fig.2 left panels (plots in $XZ$ and $YZ$ planes). However as V increases the amplitudes become ordered and move towards a fixed value tending to limit cycle oscillation at that fixed value of amplitude as is shown in Fig.2. This disorder to orderness in amplitudes as a function of $V$ could be one way to look at the transition from stochastic to deterministic system by defining an order parameter $\eta_A$ which is noise in amplitude given by, $\eta_A=\frac{\sigma_A}{\langle A\rangle}$. Here $\langle A\rangle$ is the mean of the random amplitudes of the spikes for a particular $V$ and $\sigma_A=\langle(\langle A\rangle-A)^{2}\rangle^{1/2}$ is the standard deviation of the amplitudes. The transition from stochastic to deterministic can be identified by looking at the behavior of $\eta_A$ as a function of V. At thermodynamic limit i.e. at $V\rightarrow\infty$, essentially $\eta_A\rightarrow 0$ because of the transition from disorder or randomness to orderness in $A$. Numerically one can define a critical value $V_c$ such that: (1) if $V\langle V_c$, $\eta_A$ will have values larger than a $\eta^c_A$ as $V$ decreases and so the system is in stochastic regime, (2) if $V\rangle V_c$ the values of $\eta_A$ will remain almost stationary to $\eta^c_A$ as $V$ increases, and therefore the system is in deterministic regime. 

We now simulate the oregonator model (\ref{ore}) to obtain $\eta_A$ for each value of $V$ by averaging the amplitudes of 300 spikes for each value of $V$. The plots of $\eta_A$ for $X$ and $Y$ variables are shown in the right hand panels in Fig.2. From these plots we could able to identify the approximate critical values of $V$ and $\eta_A$ to be $V_c\sim 70\pm 5$ and $\eta_c\sim 0.004\pm 0.001$ respectively. In the stochastic regime ($V\langle V_c$) $\eta_A$ increases as $V$ decreases, whereas in $V\rangle V_c$, $\eta_A$ is stationary as $V$ increases. The stochastic and deterministic regimes are shown in the $(\eta_A-V)$ phase diagrams as shown in the panels.

We now follow Gillespie's technique \cite{gil2} to reduce the Master equation (\ref{more}) to a more simpler form, the Chemical Langevin Equation (CLE), based on two important realistic approximations made on a random variable $K(\vec{x},\tau)$ which is the number of a particular reaction fired during a time interval $[t,t+\tau]$ with $\tau\rangle 0$. The first approximation is to impose small $\tau$ limit that let the propensity function remain constant ($\omega\sim constant$) during the time interval and $K$ to approximate to statistically independent Poisson distribution functions. The second approximation is to impose large $\tau$ limit that let $\omega\rangle\rangle 1$ and Poisson distribution function to be replaced by Normal distribution function with same mean and variance. Following these steps we reach the following CLE for oregonator,
\begin{widetext}
\begin{eqnarray}
\label{cle}
\frac{dX}{dt}&=&k_1AY-k_2XY+k_3CX-k_4X^2+\frac{1}{\sqrt{V}}\left[\sqrt{k_1AY}\xi_1-\sqrt{k_2XY}\xi_2+\sqrt{k_3CX}\xi_3-\sqrt{k_4X^2}\xi_4\right]\nonumber\\
\frac{dY}{dt}&=&-k_1AY-k_2XY+k_5EZ+\frac{1}{\sqrt{V}}\left[-\sqrt{k_1AY}\xi_5-\sqrt{k_2XY}\xi_6+\sqrt{k_5EZ}\xi_7\right]\\
\frac{dZ}{dt}&=&k_3CX-k_5EZ+\frac{1}{\sqrt{V}}\left[\sqrt{k_3CX}\xi_8-\sqrt{k_5EZ}\xi_9\right]\nonumber
\end{eqnarray}
\end{widetext} 
where, $\xi_i=lim_{dt\rightarrow 0}N_i(0,1)/\sqrt{dt}$ are noise parameters satisfying, $\xi_i(t)\xi_j(t^\prime)=\delta_{ij}\delta(t-t^\prime)$. $\{N(0,1)\}$ are normal distribution functions with mean 0 and variance 1. The noise terms in the dynamics of each variables are functions of $V$, $\xi$ and the variables in the system. If the noise term is negligible, the equation (\ref{cle}) reduces to deterministic equations whose steady state solutions can be obtained by putting $\frac{d}{dt}\vec{x}(t)=0$ which are given by,
\begin{eqnarray}
X^*_d&=&\frac{k_1A}{2k_2}\left(\sqrt{1+\frac{8Ck_2k_3}{Ak_1k_4}}-1\right)\nonumber\\
Y^*_d&=&\frac{k_1k_4}{4k_2^2}\left(\frac{4Ck_2k_3}{Ak_1k_4}-\sqrt{1+\frac{8Ck_2k_3}{Ak_1k_4}}\right)\\
Z^*_d&=&\frac{k_1k_3AC}{2k_2k_5E}\left(\sqrt{1+\frac{8Ck_2k_3}{Ak_1k_4}}-1\right)\nonumber
\end{eqnarray}
If we represent $\vec{x}_{c}(t)$ and $\vec{x}_{d}(t)$ as the variables obtained by solving CLE in (\ref{cle}) and deterministic equations, then the noise term associated with variables $\vec{x}_{c}(t)$ can be estimated by $\eta_L(t)\sim [\vec{x}_{c}(t)-\vec{x}_{d}(t)]$. From equation (\ref{cle}) we can write the noise term as,
\begin{eqnarray}
\label{noise}
\eta_L(\vec{x}_d,\vec{\xi},V;t)&=&\int\left[\vec{x}_{c}(t)-\vec{x}_{d}(t)\right]dt\nonumber\\
&=&\frac{1}{\sqrt{V}}\int F\left(\vec{x}_d,\vec{\xi},V;t\right)dt
\end{eqnarray}
where $F(\vec{x}_d,\vec{\xi},V;t)$ is a function and $\vec{\xi}$=$[\xi_1,\xi_2,\cdots,\xi_N]^{-1}$, $N=9$ in the case of oregonator model we consider. This noise parameter can also be used to detect stochastic to deterministic transition in the same fashion we have done in the case of stochastic simulation. Now we solve CLE (\ref{cle}) using standard Runge-Kutta method for numerical integration of a set of differential equations \cite{pre} for different values of $V$. For each $V$ we calculate noise parameter in amplitude $\eta_L$ as we have done above in the case of SSA by taking 300 spikes. The results are shown in Fig.3. The results show that $\eta_L^X (\eta_L^Y)\rangle \eta_A (X,Y)$ in magnitudes. This shows that CLE system described by (\ref{cle}) is much much closer to deterministic system as compared to ME system given by (\ref{more}).

Since there is no exact and clear demarcation between actual stochastic and deterministic systems, the attempt to detect the transition between these two regimes with exact number is almost not possible. However we can look for some order parameter which can able to detect this transition approximately. For example we used noise parameter, which is calculated from random amplitudes as a function of system size, as order parameter to detect this transition. The randomness in the amplitudes of each individual spikes in the dynamics of the variables involved in a stochastic system in a particular system size is due to noise associated with these variables induced by random particle interaction in the system. This randomness in amplitudes will become order and tends to a fixed value of amplitude as the stochastic system ($V_s$) goes to deterministic limit ($V_d$) i.e as $(V_s\rightarrow V_d)$. This leads to the transition from the fluctuating limit cycle oscillations to the fixed limit cycle oscillation. This transition can be well characterized by the noise parameter in amplitudes $(\eta_A,~\eta_L)$ and we use it as an order parameter to detect the transition. We could able to detect the transition from the behavior of $\eta_A$ which shows increasing nature as $V$ decreases and minimum stationary value as $V$ increases. However there are some issues in our numerical experiments, for example it is very difficult to find ideal deterministic limit where $\eta_A$ should be zero. Further it is hard to decide the number of spikes to be taken for calculating $\langle A\rangle$ (we took 300 spikes) so that the value of $\eta_A$ to be obtained is statistically significant and correct. 

The Master equation formalism of the chemical oscillator model we have taken and the SSA simulation systematically takes into account the noise fluctuations in the dynamics of the variables from the particle interaction picture. However, the Chemical Langevin Equation beautifully can able to connect stochastic and deterministic descriptions scaling by volume and the noise terms fluctuating with the order of $V^{-1/2}$ letting the variables evolve with time. But the value of $\eta_L^X$ and $\eta_L^Y$ obtained in CLE is found to be much smaller than the value $\eta_A$ found in ME.

Most living systems probably use noise fluctuations in a more constructive way, within an optitimal level, in weak signal amplification and information processing. However if the strength of the noise is larger than this optimal value then the role of the noise become in the destruction of the signal. It is also to be noted that the size of living cells from birth to death also fluctuate in short time scales. In such situations the role of noise could be different from time to time whenever changes in the cellular size takes place. Such changes in the role of noise could affect on various biological functions, cellular communications, signal manipulations etc. The investigation of such problems could explore important issues regarding how biological, chemical and physical systems work.

{\large\bf Acknowledgment:} We would like to thank R. Ramaswamy for his encouragement useful discussion to carry out this work. This work is financially supported by Department of Science and Technology (DST) and carried out in the Center for Interdesciplinary Research in Basic Sciences, Jamia Millia Islamia, New Delhi,India.


\begin{thebibliography}{99}

\bibitem{rao}Rao CV, Wolf DM and Arkin AP (2002): {\it Nature (London) 420}: 231.
\bibitem{swa} Swain PS, Elowitz MB and Siggia ED (2002): {\it PNAS 99}: 12795-12800.
\bibitem{gil1} Gillespie DT (1977): {\it J. Phys. Chem. 81}: 2340-2361.
\bibitem{gil2} Gillespie DT (2000): {\it J. Chem. Phys. 113}: 297-306.
\bibitem{bla} Blake WJ, Kaern M, Cantor CR and Collins JJ (2003), {\it Nature (London) 422}: 633.
\bibitem{sco} Scott M, Ingalls B and Kaern M (2006): {\it Chaos 16}: 026107.
\bibitem{mcq} McQuiarre DA (1967): {\it J. Appl. Probab. 4}: 413.
\bibitem{han} Hanggi P (2002): {\it ChemPhysChem 3}: 285-290.
\bibitem{ani} Anishchenko V.S., Neiman A.B., Moss F and Schimansky-Geier L (1990): {\it Physcs-Uspekhi 42}: 7-36.
\bibitem{bas} Bassler BL (1999): {\it Curr. Opin. Microbiol. 2}: 582-587.
\bibitem{gil3} Gillespie DT J. Phys. Chem. B 113 (2009) 1640-1644.
\bibitem{sco1} Scott M, T. Hwa and Ingalls B., Proc. Nat. Acad. Sc. 104 (2007), 7402-7407.
\bibitem{fie} R. J. Field and R. M. Noyes, J. Chem. Phys., 60, 1877 (1973
\bibitem{tys} Tyson J.J., J. Chem. Phys. 58, (1973) 3919. Tyson J.J. and Light J.C. J. Chem. Phys. 59, (1973) 4164.
\bibitem{pre} W.H. Press, S.A. Teukolsky, W.T. Vetterling and B.P. Flannery, Numerical Recipe in Fortran, Cambridge University Press, 1992. 

\end{thebibliography}
\end{document}